\documentstyle[12pt,aaspp]{article}

\received{1993 October 6}
\revised{DATE}
\accepted{1993 December 7}
\journalid{VOL}{JOURNAL DATE}
\articleid{Start}{Finish}
\paperid{Manuscript id}
\lefthead{BAND}
\righthead{EFFECT OF REPEATERS ON V/V$_{max}$}

\def\vvmax{\hbox{V/V$_{max}$}}
\def\vvmaxav{\hbox{$\langle$V/V$_{max} \rangle$}}
\def\ccmax{\hbox{C$_{max}$/C$_{min}$}}

\begin{document}
\title{THE EFFECT OF REPEATING\\GAMMA RAY BURSTS ON \vvmax}
\author{David L. Band}
\affil{Center for Astrophysics and Space Sciences \\
University of California, San Diego, CA  92093-0111 \\
I: dlbbat@cass09.ucsd.edu \\
{\it Received 1993 October 6; accepted 1993 December 7}}
\begin{abstract}
I consider the effect of repeating burst sources on the \vvmaxav\ statistic.
I find that if the burst sources are distributed homogeneously in a
$d$-dimensional space, \vvmaxav\ converges to $d/(d+3)$ as long as the
luminosity function is independent of position.  Choosing the brightest event
from a cluster of $k$ events shifts the luminosity function to higher
luminosities, but if the original luminosity function is position-independent,
it remains so.  Therefore the treatment of repeating events, if applied
consistently, will not affect the effectiveness of \vvmaxav\ as a test of
burst homogeneity.  The calculation of \vvmaxav\ for apparent repeating and
nonrepeating source populations will be biased by the incorrect
classification of faint bursts.  In conclusion, the current practice of
calculating \vvmaxav\ using all bursts, even apparent repeaters, and
treating multi-spike bursts as single bursts, is valid and consistent.
\end{abstract}
\section{Introduction}
Before the release of the first gamma ray burst catalog (Fishman
et al. 1993) from the Burst and Transient
Source Experiment (BATSE) on the Compton Gamma Ray Observatory
(GRO), the traditional view had been that the ``classical'' burst sources
did not repeat while the three known repeating gamma ray burst sources
constituted a different class, the soft gamma-ray burst repeaters (SGRs),
with different spectral properties (see e.g., Higdon \& Lingenfelter 1990).
Recent studies of this catalog suggest that the classical bursts do
indeed repeat:  Wang \& Lingenfelter (1993a,b) have identified a
repeating source; and Quashnock \& Lamb (1993) find an excess
of nearest-neighbors at small separation, suggesting a population of
repeaters.  However, Hartmann et al. (1993) do not find the expected
signatures for repeaters in the angular correlation function or the
nearest-neighbor statistic.  Similarly Narayan \& Piran (1993) find only
a weak nearest-neighbor excess at small separation, and a comparable
excess of antipodal bursts (i.e., farthest neighbors with separations
close to 180$^\circ$), which they find symptomatic of systematic
effects and not of repeaters; Maoz (1993) discusses a possible systematic
effect.  Nowak (1993) identifies weaknesses in
the nearest-neighbor methodology.  Therefore, the existence of repeaters
is controversial.  In addition, Lingenfelter, Wang \& Higdon
(1993) suggest that complex, multi-spike bursts should be treated as
a series of repeating events.

In light of the possibility of repeating burst sources, and the suggestion
that each spike within a burst should be considered as a burst in its own
right, the effect of repeating bursts on the \vvmaxav\ test for source density
uniformity (Schmidt 1968; Schmidt, Higdon \& Hueter 1988) needs to be
evaluated.  In this study I use ``homogeneous'' as uniform in flat Euclidean
space.  I consider a population of events drawn from the same luminosity
function without regard to the prior history (i.e., each event is independent).
Enough sources are represented so that space is sampled sufficiently.
Throughout I deal with the effects of the distance to the burst source and not
the projected position on the sky.  The events from each source are clustered,
and \vvmaxav\ can be calculated either by including all events, or only the
brightest.

The cluster of events might be repeated bursts from the same source (i.e.,
events separated by a few hours); since rarely can we identify repetitions,
\vvmaxav\ is usually calculated using each event.  The assumption that each
burst is independent (i.e., characterized by a single luminosity function) is
a reasonable assumption in the absence of further information.  On the other
hand, a cluster of events within a few minutes, which are undoubtedly from
one source given the overall burst frequency, is currently considered to be
one burst, and \vvmaxav\ is calculated using the brightest spike within the
cluster.  The hard-to-soft evolution seen in many bursts (Norris et al. 1986;
Band et al. 1992) shows that the burst source retains memory of preceding
spikes.  Therefore the assumption that the events are drawn independently
from a single luminosity function is mediocre.  Nonetheless, I assume a
single luminosity function for the purpose of the current analysis.

My argument is as follows:  after developing the formalism for calculating
\vvmaxav\ (\S 2), I show that \vvmaxav\ is independent of source luminosity
functions
which are functions of luminosity alone (i.e., are not distance-dependent)
if the source population is uniform in some dimension (e.g., a sphere
or a disk).  Considering all the events from a repeater, or retaining
only the brightest, affects the luminosity function but maintains its
spatial independence (\S 3).  Separating bursts into apparent repeaters and
nonrepeaters introduces spatial-dependence into the luminosity function,
with \vvmaxav\ biased to lower values for apparent repeaters, and to higher
values for nonrepeaters (\S 4).
Finally, I draw the relevant operational conclusions (\S 5).
\section{Functional Dependence of V/V$_{max}$}
Assume bursts have the normalized differential luminosity distribution
$\phi[L,r]$ where the peak photon luminosity $L$ is in photons-sec$^{-1}$,
and the burst sources are distributed as $n[r]$.  Note that I permit the
luminosity function to change with radius.  The observed peak count rate
is $C_{max}=L/4\pi r^2$. At any given time the detector has a count rate
threshold $C_{min}$ with a normalized distribution $g[C_{min}]$.
Define $\xi=$\ccmax\ and $v=\xi^{-3/2}$ (this is the transformation
that removes the detector threshold to answer specific questions; see
Band 1993a).  Then the distribution of $\xi$, $\Xi[\xi]=dN/d\xi$, is
\begin{eqnarray}
 \Xi[\xi] &=& \int n[r] g[C_{min}] \phi[L,r]
   \delta \left[C_{max}- {L \over{4\pi r^2}}\right]
   \delta \left[\xi- {{C_{max}}\over{C_{min}}}\right]
   \,dL \,dV \, dC_{min} \, dC_{max} \\
&=& \int n\left[ \left( {L\over{4\pi \xi C_{min}}}\right)^{1/2}\right]
   \phi \left[ L,\left( {L\over{4\pi \xi C_{min}}}\right)^{1/2}\right]
   {{g[C_{min}] \xi^{-5/2}}\over{4\sqrt{\pi}}} \left({L \over {C_{min}}}
   \right)^{3/2} \,dL \, dC_{min} \quad . \nonumber
\end{eqnarray}
In brief, the integral in eqn. (1) considers all the sources which
can contribute bursts with a value $\xi$:  as the threshold $C_{min}$
varies (described by $g[C_{min} ]$), different values of $C_{max}$
contribute to $\Xi$; a given $C_{max}$ fixes the ratio of $L$ to $r^2$; the
luminosity function $\phi[L,r]$ therefore weights the distribution of relevant
radii; and finally, the source density $n[r]$ provides the number of
burst sources at each radius.
To get the average \vvmax=$v=\xi^{-3/2}$ I need the distribution of $v$
\begin{eqnarray}
\chi[v] &=& \Xi [\xi] {{d\xi}\over{dv}} = -{2\over3} v^{-5/3} \Xi [\xi] \\
   &=& {1\over{6\sqrt{\pi}}} \int
   n\left[ \left( {{L v^{2/3}} \over{4\pi C_{min}}}\right)^{1/2}\right]
   \phi \left[ L,\left( {{L v^{2/3}} \over{4\pi C_{min}}}\right)^{1/2}\right]
   g[C_{min}] \left({L \over{C_{min}}}\right)^{3/2} \,dL \, dC_{min} \nonumber
\end{eqnarray}
where I use $dN = \Xi d\xi = \chi dv$ (and I drop the minus
sign).  The average \vvmax\ is
\begin{equation}
\vvmaxav = \int_0^1 dv \,v\,\chi[v]  / \int_0^1 dv \,\chi[v]
   \quad .
\end{equation}

Assume $n(r)=n_0 (r/r_0)^\alpha$; clearly the homogeneous 3-dimensional case
corresponds to $\alpha=0$.  Then (inserting $n(r)$ into eqn. [2])
\begin{equation}
\chi[v] = {{n_0 v^{\alpha/3}}\over{3 (4\pi)^{(\alpha+1)/2} r_0^\alpha}}
   \int \left({L \over {C_{min}}}\right) ^{{3+\alpha} \over 2}
   \phi \left[ L,\left( {{L v^{2/3}} \over{4\pi C_{min}}}\right)^{1/2}\right]
   g[C_{min}] \,dL \, dC_{min} \quad .
\end{equation}
If the luminosity function $\phi[L,r]$ is a function of luminosity
alone, $\phi[L]$, and has no spatial dependence then the dependence
of $\chi[v]$ on $v$=\vvmax\ is decoupled from the luminosity function
(i.e., in eqn. [4] the integrand has no $v$ dependence).  In this case
\begin{equation}
\vvmaxav = {{\int_0^1 dv \,v^{(\alpha+3)/3}}\over{\int_0^1 dv \,v^{\alpha/3}}}
   = {{\alpha+3}\over {\alpha+6}}   \quad .
\end{equation}
For source distributions homogeneous in a $d$-dimensional space
$n(r) dV \propto r^{d-1} dr$.  The equivalent 3-dimensional source
density which can be used in the above equations for $\chi[v]$ is
$n(r) \propto r^{d-3} $ or $\alpha=d-3$; this is the density the
spatial distribution would have if averaged spherically.  Consequently, if
the luminosity function is spatially independent then
\begin{equation}
\vvmaxav = {d \over {d+3}}
\end{equation}
which gives \vvmaxav=0.5, 0.4 and 0.25 for $d=$3 (sphere), 2 (disk---Galactic
plane?) and 1 (linear---spiral arm?).  An important corollary is that any
manipulation that maintains the spatial independence of the luminosity
function will not affect \vvmaxav.

In addition to its average values for homogeneous source densities,
\vvmax\ will be distributed uniformly between 0 and 1 for $d=3$, providing
a secondary test for this dimension.  Variants tailored to dimensions
other than $d=3$, such as the $\langle A/A_{max} \rangle$ test for $d=2$
(Kluzniak 1992), can be constructed by transforming to $\xi^{-d/2}$.
\section{The Luminosity Function of Burst Clusters}
Assume repeating sources all have the same spatially independent luminosity
function $\phi_1(L)$, and that the source sample is large enough so that true
averages are observed.  The repeating sources can be treated either by
considering only the brightest event in calculating \vvmaxav, or by
including all observed events.  I will show that the luminosity function
for each case is spatially independent, and will find the effect these
choices have on \vvmaxav.

First I find the luminosity function $\phi_k[L]$ for the brightest event
in a cluster of $k$ events. Since the differential luminosity function is
normalized, the cumulative luminosity function $\Phi_1[L]$ varies uniformly
between 0 and 1 (by definition).  Therefore this cumulative distribution
is a mapping of the luminosity into a uniform distribution:
\begin{equation}
g_1[u] = 1 \hbox{ for } 0\le u \le 1 \quad \hbox{where } \quad
   u = \int_0^L \phi_1[L^\prime] dL^\prime  =\Phi_1[L] \quad .
\end{equation}
If there are $k$ events in a cluster, then the distribution of the value
of the single event cumulative luminosity function corresponding
to the brightest event is
\begin{equation}
g_k[u] = k u^{k-1} \quad \hbox{ for } \quad 0\le u \le 1
\end{equation}
because there are $k$ events which could be the brightest and there are $k-1$
other events which must be less bright; the probability that an event is less
bright is $u$.  Note that since the average value of $u$ is
\begin{equation}
\langle u \rangle = \int_0^1 g_k(u) \, u \,du = {k\over{k+1}} \quad ,
\end{equation}
the average luminosity is shifted to higher values, as expected.  The increase
is not very great for flat luminosity functions (e.g., by no more than a
factor of 4 for $\phi_1\propto L^{-1/2}$), but can be very dramatic for steep
luminosity functions (e.g., by more than a factor of 50 for a ten-event
cluster and $\phi_1\propto L^{-2}$).  The
luminosity function of the brightest burst is derived by relating it to the
single event luminosity function.  The cumulative luminosity
function of the brightest event $\Phi_k$ is also distributed uniformly:
\begin{equation}
G[U] = 1 \quad \hbox{ for } \quad 0\le U \le 1 \quad \hbox{where } \quad
   U = \int_0^L \phi_k[L^\prime] dL^\prime =\Phi_k[L] \quad .
\end{equation}
Equating $G[U] dU = g_k [u]du$ gives
\begin{eqnarray}
U &=& u^k \nonumber \\
\int_0^L \phi_k[L^\prime] dL^\prime &=&
   \left( \int_0^L \phi_1[L^\prime] dL^\prime\right)^k \\
\phi_k[L] &=& k \Phi_1[L]^{k-1} \phi_1[L] \quad . \nonumber
\end{eqnarray}
Note that $\phi_k[L]$ is the luminosity function regardless of whether
all $k$ events can be detected.  Similarly, since each event is assumed
to be independent, a cluster of $M$ events can be considered as two
clusters of $N$ and $M-N$ events if the separation into two clusters
is based on a luminosity-independent criterion, e.g., the temporal
separation between events.

Let $f_k$ be the fraction of clusters with $k$ events; obviously
$\sum_{k=1}^\infty f_k$ = 1, and the average number of events
$\sum_{k=1}^\infty k\,f_k$ must converge.  If each event is considered
separately,
\begin{equation}
\phi[L] = \phi_1[L] \sum_{k=1}^\infty k\,f_k
\end{equation}
where the average number of events has been inserted to preserve
the total number of events observed ($n[r]$ is the density of sources).
However, if we consider only the brightest event in each cluster, then
\begin{equation}
\phi[L] = \sum_{k=1}^\infty f_k \phi_k[L] \quad .
\end{equation}
Eqn. (13) is applicable if the events from a repeating source are
broken into a number of clusters based on a luminosity-independent
criteria.  For example, events separated by less than some predetermined
time (e.g., half an hour) may be considered a single burst, events after
a gap greater than this time are considered as part of a new burst.
\section{Separating Apparent Repeaters and Nonrepeaters}
There is a temptation to separate repeating burst sources from nonrepeaters,
and calculate \vvmaxav\ for each class, perhaps to determine
whether they originate from different source populations.  However,
repeaters can only be identified if the second brightest event exceeds
the detector threshold.  Whether the repeating source is identified as
such therefore depends on the distance to the source.  This introduces a
spatial dependence into the luminosity functions, complicating the
determination of the source distribution.  Sources observed to repeat
will be closer, and have a smaller \vvmaxav.

Consider a cluster of $k$
events with the luminosity of the brightest event corresponding to a value
$u_1$ of the single event cumulative luminosity function (see eqn. [7]).
Then $u_2$, the value of the single event cumulative distribution for
the second brightest burst, is distributed as
\begin{equation}
g_{k,2}[u_2 \,|\, u_1] = (k-1)u_2^{k-2} u_1^{1-k} , \qquad u_2<u_1 \quad ;
\end{equation}
there are $k-1$ possible second brightest events, and the $k-2$ less bright
events must have values of $u$ less than $u_2$.  The dependence on $u_1$
results from restricting all $k-1$ bursts to the range 0 to $u_1$.  Then the
fraction of the clusters for which the brightest event has a value $u_1$ and
the second event has a luminosity which produces a count rate less than
threshold (i.e., $L_2<L_0=4\pi r^2 C_{min}$) is
\begin{eqnarray}
f_{s,k} &=& \int_0^{u_0} (k-1) u_2^{k-2} u_1^{1-k} du_2 =(u_0 /u_1)^{k-1}
   \nonumber \\
   &=& \left( {{\int_0^{L_0} \phi_1[L^\prime] dL^\prime }
   \over{\int_0^{L} \phi_1[L^\prime] dL^\prime}} \right)^{k-1}
= \left( {{\int_0^{L v^{2/3}} \phi_1[L^\prime] dL^\prime }
   \over{\int_0^{L} \phi_1[L^\prime] dL^\prime}} \right)^{k-1}
\end{eqnarray}
where $u_0=\Phi_1[L_0]$ and
$L_0 = 4\pi r^2 C_{min}= 4\pi r^2 C_{max} \xi^{-1} = L v^{2/3} $.
Thus $f_{s,k}$ is the fraction of the $k$-event clusters which are counted
as nonrepeaters (i.e., misidentified) because the second spike is
below threshold.  Note that $f_{s,k}$ is a function of both the intensity
of the brightest event, and the distance to the source (through $L_0$).
The effective differential luminosity functions for apparent single events
$\phi_s$ and for apparent multiple events $\phi_m$ are
\begin{equation}
\phi_s = f_1\phi_1 + \sum_{k=2}^\infty f_{s,k} f_k \phi_k
   \quad \hbox{and} \quad
\phi_m = \sum_{k=2}^\infty (1-f_{s,k}) f_k \phi_k \quad .
\end{equation}
Because $f_{s,k}$ is a function of distance to the burst source, the
luminosity function for
the apparent repeaters and nonrepeaters is also spatially-dependent.
Since $f_{s,k}$ increases with $v$, it causes $k$-event clusters to be
misclassified as apparently single events with larger values of
\vvmaxav\ and correctly classified as repeaters with smaller \vvmaxav.
\section{Conclusions}
The \vvmaxav\ statistic is best used to test whether sources are distributed
uniformly in Euclidean, 3-dimensional space (Band 1993a), or by extension, in
any other $d$-dimensional space.  I showed that this diagnostic power assumes
the luminosity function is spatially independent.  Therefore, transformations
which do not introduce a spatial dependence into the luminosity function will
not affect the utility of \vvmaxav.

Using only the brightest event from a cluster of $k$ events modifies the
effective luminosity function but maintains its spatial independence.  As
expected, the average luminosity increases with the number of events, which
means that \vvmaxav\ probes greater distances, and is therefore more sensitive
to inhomogeneities in the source distribution.  However, if the source
distribution is indeed homogeneous, the treatment of event clusters does not
affect \vvmaxav.

Apparent repeater and nonrepeater populations do not have meaningful values
of \vvmaxav; \vvmaxav\ is biased towards smaller values for apparent repeaters
and towards larger values for apparent nonrepeaters.  The second brightest
event from a repeater must be detectable, and therefore the fraction of true
repeaters which are classified as nonrepeaters increases with distance.  Thus
multi-spiked bursts should not be separated from single-spike bursts, and
repeating sources from nonrepeating.

The burst phenomenon may consist of events with a distribution of separation
times peaked at short timescales, but with a tail to large separations
(Lingenfelter et al. 1993).  Those events within a few minutes of each other
are treated as a single burst, while those separated by hours or more
are considered different bursts.  Note that the existence of repeating
bursts separated by more than a few minutes is controversial, as discussed
in the Introduction.  This separation into individual multi-spike bursts and
repeating bursts does not depend on the distance to the burst source, and
should not introduce a spatial dependence into the luminosity function.
Therefore the current method of calculating \vvmaxav\ does not compromise
the test for homogeneity.

There are two alternatives to the current methodology:  calculating
\vvmaxav\ using each spike in a burst or including only the strongest
burst from a repeating source.  If performed consistently, each
alternative will not affect the diagnostic power of \vvmaxav.  However,
the assumption that in a multi-spike burst each spike is drawn independently
from the same luminosity function is invalidated
by the observed spectral evolution.
Similarly, it is probably impossible to identify distant repeating sources
or faint repetitions from nearby sources as a consequence of the large
uncertainty in the projected position on the sky of low count rate bursts;
incomplete removal of all but the
strongest event from a repeater will bias \vvmaxav\ to higher values.
Therefore these variants do not increase the effectiveness of \vvmaxav,
and can easily introduce biases.

Previously I (and others) showed that \vvmaxav\ is appropriate for testing
the hypothesis of source homogeneity, but not for investigating inhomogeneous
source distributions (Hartmann \& The 1992; Petrosian 1993; Band 1993a).
Maximum-likelihood methods using the observed peak flux and the known
detection threshold distributions (Band 1993b; Loredo \& Wasserman 1993),
survival analysis (Efron \& Petrosian 1993), and moment methods (Horack et al.
1993) are more powerful for studying inhomogeneous sources.  Since BATSE has
found that \vvmaxav\ is definitely not 1/2, and the intensity distribution is
not a power law (e.g., Meegan et al. 1993; Fishman et al. 1993), the burst
source distribution is clearly not homogeneous.  Therefore, \vvmaxav\ should
be retired in favor of
methods which utilize all the information in the burst distribution.  For these
superior methods the treatment of repeaters discussed here is just as relevant.
Since the luminosity function is unknown and must be modeled, manipulations
which alter the luminosity function will not complicate significantly an
already difficult problem.  However, it is preferable to keep the luminosity
function distance-independent.

In summary, the current practice of treating all events which are separated
by less than a reasonable time (e.g., $\sim$half-an-hour) as one burst and
retaining all the bursts from an apparent repeater produces a burst database
appropriate both for the calculation of \vvmaxav\ and for studies of
inhomogeneous source models.
\acknowledgements
It is a pleasure to thank M.~Briggs, L.~Ford, D.~Gruber, R.~Lingenfelter,
R.~Rothschild, and V.~Wang for incisive discussions and assistance in
improving this text.  This work was supported by NASA contract NAS8-36081.

\end{document}